# High-resolution Power Doppler Using Null Subtraction Imaging

Zhengchang Kou, *Member, IEEE,* Matthew Lowerison, Qi You, Yike Wang, *Student Member, IEEE,*

Pengfei Song, *Senior Member, IEEE* and Michael L. Oelze, *Senior Member, IEEE*

*Abstract*— **To improve the spatial resolution of power Doppler (PD) imaging, we explored null subtraction imaging (NSI) as an alternative beamforming technique to delay-and-sum (DAS). NSI is a nonlinear beamforming approach that uses three different apodizations on receive and incoherently sums the beamformed envelopes. NSI uses a null in the beam pattern to improve the lateral resolution, which we apply here for improving PD spatial resolution both with and without contrast microbubbles. In this study, we used NSI with three types of singular value decomposition (SVD)-based clutter filters and noise equalization to generate high-resolution PD images. An element sensitivity correction scheme was also proposed as a crucial component of NSI-based PD imaging. First, a microbubble trace experiment was performed to evaluate the resolution improvement of NSI-based PD over traditional DAS-based PD. Then, both contrast-enhanced and contrast free ultrasound PD images were generated from the scan of a rat brain. The cross-sectional profile of the microbubble traces and microvessels were plotted. FWHM was also estimated to provide a quantitative metric. Furthermore, iso-frequency curves were calculated to provide a resolution evaluation metric over the global field of view. Up to six-fold resolution improvement was demonstrated by the FWHM estimate and four-fold resolution improvement was demonstrated by the iso-frequency curve from the NSI-based PD microvessel images compared to microvessel images generated by traditional DAS-based beamforming. A resolvability of 39 μm was measured from the NSI-based PD microvessel image. The computational cost of NSI-based PD was only increased by 40 percent over the DAS-based PD.**

## I. INTRODUCTION

High-sensitivity power Doppler (PD) imaging [1][2][3] has emerged as a powerful method for visualizing tissue microvasculature in a variety of applications. High-sensitivity PD uses ultrafast imaging (e.g., several thousands of Hertz of frame rate) to boost Doppler sensitivity [4] to blood flow in small vessels. Extensive efforts have been conducted to improve the performance of ultrafast PD, which include developing advanced clutter filters [5], beamforming [6], compounding [7] and denoising filters [8].

In terms of exploring advanced clutter filters, singular value decomposition (SVD) has been widely adopted for high sensitivity PD imaging [9]. To improve the clutter rejection performance, Song *et al* proposed a block-wise adaptive local SVD filter [5], which significantly improved the image quality. Ozgun *et al* proposed to use a higher order SVD [10] to the delay compensated aperture data, which demonstrated improvement in contrast-to-noise ratio (CNR) as compared to a global SVD filter. As an extension to the SVD clutter filter, a cost-effective noise equalization method based on the smallest singular value was proposed by Song *et al* [11]. This method improves the visibility of microvessels across different depths.

Coherent plane wave compounding (CPWC) [12] is also widely used for ultrafast PD microvessel imaging [13] because it provides better signal-to-noise ratio (SNR) and CNR performance over single plane wave imaging along with a larger field of view. Different beamforming and compounding methods have been explored based on CPWC to further improve image quality. Stanziola *et al* proposed acoustic subaperture processing (ASAP) [6], which splits the channel data into two subgroups and averages the correlation results of the beamformed data from two subgroups to get the final image. As noise is uncorrelated between subgroups, ASAP could substantially improve SNR performance. Huang *et al* proposed a method similar to ASAP by splitting tilted planes or diverging waves into two subgroups and performing coherent compounding on each subgroup independently [14]. Jakovljevic *et al* proposed short-lag angular coherence [7], which utilizes the angular coherence between different plane-wave transmits to suppress the incoherent noise and motion artifacts. Kang *et al* proposed frame-multiply-and-sum based ultrafast PD imaging [15] by utilizing the coherence between plane-wave angle frames to improve both CNR and SNR of the image. Shen *et al* proposed to use delay-multiply-and-sum-based beamforming with complementary subset transmit to improve SNR [16]. Shen *et al* proposed to use temporal

The work of Zhengchang Kou was supported by Beckman Institute Graduate Fellowship and Beckman Institute Postdoctoral Fellowship. The work of Matthew R. Lowerison was supported by Beckman Institute Postdoctoral Fellowship. This work was supported in part by the by the National Institutes of Health under grants R21EB024133, R21EB023403, R21EB030743, R01CA251939, R21EB030072, R01EB031040, and R21AG077173, and in part by the National Science Foundation under Award 2237166. (Corresponding author: Michael L. Oelze).

All authors are affiliated with the Beckman Institute for Advanced Science and Technology, University of Illinois at Urbana-Champaign, Urbana, IL 61820 USA (email: zkou2@illinois.com; oelze@illinois.edu); In addition, P. Song and M.L. Oelze are with the Department of Electrical and Computer Engineering, the Carle Illinois College of Medicine and Department of Bioengineering.

multiply-and-sum power Doppler detection method to further improve the SNR[17].

Aside from novel coherence-based beamforming and compounding techniques, an advanced denoising filter named nonlocal means (NLM) filter [18] has also been explored as a means of reducing noise for ultrafast PD imaging. Huang et al proposed to use NLM filters on the spatiotemporal domain of clutter filtered blood flow RF data (St-NLM) to improve ultrafast PD image quality [8]. Lok et al proposed a resolution improving method for PD microvessel imaging basing on deconvolution with total variation regularization [19].

Basing on ultrafast ultrasound imaging, ultrasound localization microscopy (ULM) [22][23] has been proposed and extensively developed in the past few years. By localizing the position of microbubbles and tracking their movement, super-resolution microvessel imaging using ultrasound has been achieved by ULM. However, ULM has several challenges such as much longer acquisition time compared to ultrafast PD, the requirement for contrast agent injection and much higher computational cost. The computational cost is not only a problem for ULM, but also a limiting factor for the methods discussed previously using block-wise adaptive SVD filter [5] or NLM filters [8]. Several previous studies [7][20][21] have already discussed the importance of the computational cost and proposed methods that are computationally efficient.

To improve image quality with short acquisition time and low computational cost, we propose a beamforming-based resolution enhancement ultrafast PD using a novel computationally inexpensive non-linear beamforming technique, called null-subtraction-imaging (NSI) [24] in this study. The principle of NSI is to image with a null in the beampattern rather than the mainlobe, which is diffraction limited. By imaging with a null in the beam pattern, which is created by a Heaviside apodization, NSI drastically improves the lateral resolution beyond the diffraction limit and suppresses side lobes. The resolution improvement is still limited by the input signal SNR and a key parameter in NSI is the DC offset, which controls the mainlobe width of NSI. Recently, Kou et al demonstrated that NSI could also reduce grating lobes, which can significantly suppress image artifacts [25]. An initial study using NSI on ultrafast PD microvessel imaging demonstrated NSI's potential for detecting microvessels with contrast agents [26]. Yociss et al proposed to use NSI in volumetric contrast-enhanced vessel imaging to improve spatial resolution [27]. Inspired by NSI, Kim et al proposed the microbubble uncoupling via transmit excitation to uncouple microbubbles, which demonstrated improvement in the microbubble localization efficiency of ULM [28].

In this study, we present a framework of implementing a practical ultrafast PD microvessel imaging technique based on NSI with specially designed SVD clutter filter and noise equalization which improves the spatial resolution to overcome the diffraction limit without significantly increasing acquisition time and computational cost. We evaluate its performance *in vivo* both qualitatively and quantitatively. In Section II we describe the principle of NSI combined with an element sensitivity correction and the processing pipeline. The experimental setup and image evaluation metrics are described in Section III, followed by the results from microbubble trace experiments and animal experiments in Section IV. Discussion and conclusion of the study are provided in Sections V and VI.

## II. METHODS

### A. NSI

As the theory and performance of NSI in ultrasound imaging using a linear array have been extensively discussed and validated in previous studies [24][25], we will only offer a short description here. Briefly, NSI works by creating three apodizations for receive subapertures. One apodization is a zero mean (ZM) apodization with half the aperture having a weight of +1 and the second half with -1. This creates a sharp null at zero degrees. For the second apodization (DC1), a small DC offset is applied to the same apodization of the receive subaperture. A third apodization (DC2) is a flipped version of the second apodization with the DC offset. The means of the envelopes of the two DC offset images are subtracted by the zero mean envelope to invert the null resulting in a very sharp beam for low DC offset values.

### B. Element Sensitivity Correction

In NSI, the narrowness of the NSI beam is controlled by the DC offset parameter, i.e., a small DC offset results in a much narrower mainlobe, lower sidelobes, and lower grating lobe levels. However, when the DC offset is small, inhomogeneities of the sensitivities of individual elements become more important compared to the scenarios where the DC offset is large. This is because in NSI the formation of a null depends on having an apodization across the subaperture that is zero mean. Small differences in the sensitivities in the elements of the subaperture can prevent a true zero-mean apodization, which still results in a dip at zero degrees but not an actual null. As a result, the resolution improvement is limited. As the DC offset gets smaller, this imbalance means the DC offset can no longer function as a resolution tuning parameter for NSI. As a result, to gain resolution improvements at smaller DC offsets, it is necessary to correct for inhomogeneities in the element sensitivities. To push the limit of lowest DC offset that could still contribute to resolution improvement, we applied an element sensitivity correction (ESC) with NSI.

To perform the correction, a relative element sensitivity profile measurement for the array transducer is required. A fast measurement method of the relative sensitivity of each element of the probe is described as follows. The probe was fixed inside a water tank. One planar reflector was placed directly below the probe (Fig. 1 left shows the setup). The probe's surface and planar reflector were adjusted so that they were parallel, which was monitored via B-mode imaging to ensure the geometric delay due to the round trip was the same for all the elements. During the measurement, each element was individually excited with an imaging pulse and only the same element received the echo signal. The maximum amplitude of the received envelope resulting from the planar reflector was recorded. We defined this maximum amplitude as the transmit-receive sensitivity $S_d$, because we used the same element for both transmit and receive. The same operation was repeated on every element using the same transmit voltage.

For post-processing we only needed the receive sensitivity. Here we assumed the transmit and receive sensitivities were the same, so the single path sensitivity $S_s = \sqrt{S_d}$. To perform the ESC, we divided each receive channel data $R_{raw}(n)$ with the corresponding element's single path sensitivity $S_s(n)$, where $n$ is the receive channel number. This process is defined in Eq.1,

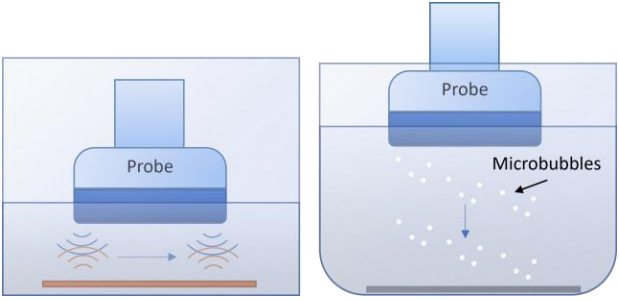

Fig. 1. Element sensitivity measurement setup (left). Each time only one element is excited with the imaging pulse and the same element receives the echo signal. The same operation is repeated for every element. Microbubble trace experiment setup (right). The array probe with microbubbles attached to the probe surface was placed in degassed water to enable the scanning of microbubbles being pushed away from the transducer due to the acoustic radiation force. Acoustic absorption material was placed at the bottom of the beaker to reduce reverberation.

$$R_{ESC}(n) = \frac{R_{raw}(n)}{S_s(n)}. \qquad (1)$$

This measurement setup is the same as an insertion loss measurement (ILM). While ILM calculates the ratio between the receive signal amplitude and the transmit signal amplitude, the proposed relative element sensitivity measurement did not calculate the ratio. Instead, it directly used the receive signal amplitude to estimate the two-way element sensitivity because the transmit voltage was the same for all elements. By using ESC, the NSI apodizations resulted in an actual weighting of each element corresponding to a smooth function closer to the desired apodization, which was essential for using small DC offsets.

*C. SVD Clutter Filter*

Normally, the SVD filter is applied on the beamformed IQ data [4]. The formation of the spatiotemporal matrix for the traditional SVD clutter filter is shown in Fig. 2. To ease the discussion, we denote this method as Post-BF-SVD as it is performed after the beamforming. The $N_f$, $N_z$ and $N_x$ represents the total number of frames, the total number of samples in axial dimension and the total number of samples in lateral dimension.

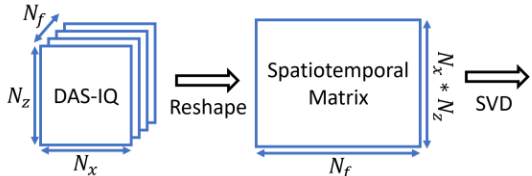

Fig. 2. Spatiotemporal matrix formation for IQ data based SVD clutter filter.

For NSI based PD, the normal SVD clutter filter is no longer adequate. This is because the subtraction of NSI is performed on the envelope data which no longer preserves the phase. To use the SVD clutter filter on NSI, we use two methods. In the first method, we apply the SVD filter on the channel data [29]. We denote this method as Pre-BF-SVD as it is performed before beamforming. Pre-BF-SVD needs to be performed on each steering angle individually. Because the size of the channel data is much larger than the IQ data and multiple SVDs need to be performed, the computational cost of Pre-BF-SVD is larger than the conventional SVD (Post-BF-SVD) clutter filter. The formation of the spatiotemporal matrix for one angle of RF data in this case is shown in Fig. 3. The $k$ represents the steering angle index.

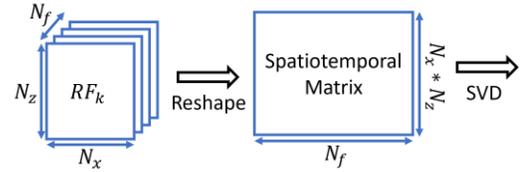

Fig. 3. Spatiotemporal matrix formation for channel data based SVD clutter filter for each steering angle.

Additionally, an SVD clutter filter that can be used with NSI after the beamforming is proposed in this study. To be specific, we concatenate each frame of the beamformed IQ data resulting from the three NSI apodizations together and reshape each concatenated frame as one column of the spatiotemporal matrix. As a result, the spatiotemporal matrix is three times larger than the conventional SVD filter that is performed on the DAS beamformed IQ data. We denote this method as Post-BF-SVD as it is performed after beamforming. The computational cost of Post-BF-SVD for NSI is higher than the traditional SVD filter for DAS because the spatiotemporal matrix is three times larger than the conventional SVD filter, but it is lower than that of the Pre-BF-SVD for NSI because Pre-BF-SVD for NSI needs to perform multiple SVDs on an even larger spatiotemporal matrix as the RF data sampling rate is normally four time the IQ data sampling rate. The formation of the spatiotemporal matrix is shown in Fig. 4.

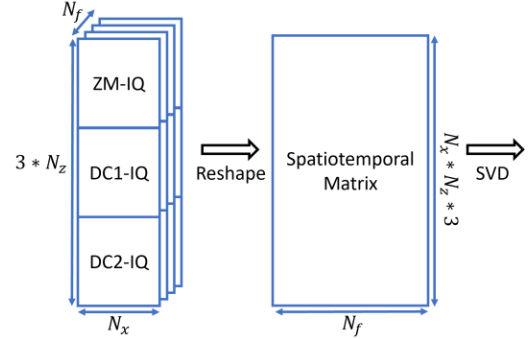

Fig. 4. Spatiotemporal matrix formation for IQ data based SVD clutter filter designed for NSI.

After formation of the spatiotemporal matrix using any of the three previously described methods, the SVD was performed using the *svd( )* function in Matlab (MathWorks, Natick, MA, USA) with economy SVD option enabled on a GPU.

To complete the comparison, we also adopted the block-wise adaptive local clutter filtering [5], which has been demonstrated as an effective method for improving the PD image quality compared with conventional global SVD clutter filter. To ease the discussion, we denote is as Blockwise-SVD in this study. A block size of 100 pixels and 90% overlap were chosen for the Blockwise-SVD. We also extend it to NSI using the spatiotemporal matrix formation methods shown in Fig. 4.

*D. Processing Chain*

In this study, PD images were generated from six different methods to complete the comparison. These six different methods are the combination of three kinds of SVD clutter filters which are Pre-BF-SVD, Post-BF-SVD, and Blockwise-SVD and two kinds of beamforming techniques which are DAS and NSI.

For Pre-BF-SVD filtered PD image, the Pre-BF-SVD filter was performed after ESC on the raw RF channel data. Then either DAS or NSI beamforming was performed before the noise equalization [11] to generate the IQ data. After noise equalization, the final PD image was generated from the IQ data with logarithm compression.

For the Post-BF-SVD and Blockwise-SVD filtered PD images, the ESC was performed on the raw RF channel data first. Then, either DAS or NSI beamforming was performed to generate the beamformed IQ data. The Post-BF-SVD filter or Blockwise-SVD filter was then applied on the IQ data.

Both the beamforming and SVD filtering were performed on an Nvidia RTX A6000 GPU (Nvidia, Santa Clara, CA, USA), which has 10752 CUDA cores and 48 GB GPU RAM using the *gpuArray( )* function of Matlab. The host computer of the GPU is a Dell Precision 5820 workstation (Dell, Round Rock, TX, USA), which has Intel Core I9 10980XE 18 cores 3.0 GHz processor (Intel, Santa Clara, CA, USA) and 256 GB DDR4 RAM. During the beamforming, the RF channel data input was first axially interpolated by a factor of 4 to improve the delay accuracy. A fixed F-number of 1 was used in the beamforming. The beamformed results were laterally interpolated by setting the target pixel lateral size smaller than the pitch during the beamforming [31]. This lateral interpolation was performed because the lower limit of the resolution depends on the output pixel size. The complete processing chain is shown in Fig. 5.

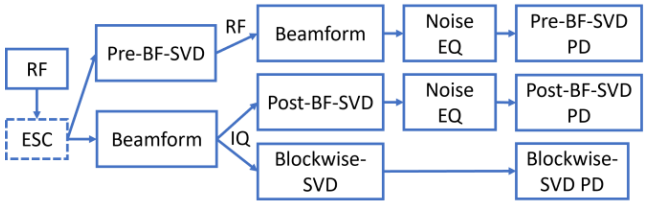

**Fig. 5.** PD microvessel imaging processing chain. The ESC is optional to evaluate its impact on image quality. The beamforming can be either DAS or NSI.

To provide a fair visual comparison, previous studies [8][21][30] manually selected the optimal dynamic range (DR). In this study, the final PD images were displayed with an automatically determined adaptive DR after subtracting the maximum value from all the pixels. This was necessary as different beamforming methods and SVD clutter filters will result in different pixel value distributions. The display dynamic range was adaptive set according to the pixel value distribution, which is defined as:

$$DR = \alpha * \sigma_{pixel} + |\mu_{pixel}| \qquad (2)$$

where $\alpha$ is task dependent. In this work, $\alpha$ was set to unity for contrast enhanced imaging and 0.5 for contrast free imaging, $\sigma_{pixel}$ is the standard deviation of the PD image pixels and $\mu_{pixel}$ is the mean of the PD image pixels. In this way, the PD image generated from different methods could be automatically scaled to provide the same level of contrast and brightness between vessel and background for the different imaging methods.

### III. EXPERIMENTAL SETUP

*A. Element Sensitivity Measurement*

As section II.B introduces, the ESC needs to be performed after the element sensitivity measurement. In this measurement, a Verasonics Vantage 256 (Verasonics, Kirkland, WA, USA) was connected to a Verasonics L22-14vX, 128-element linear array, having an 18.5 MHz center frequency to capture ultrasound data. A full cycle pulse with a center frequency of 15.625 MHz was transmitted. Due to the maximum sampling frequency limitation of the Verasonics platform, the center frequency is lower than the probe's center frequency. The probe's surface, which was under the degassed water, was 10 mm above the acrylic board sitting above the acoustic absorption material. The acrylic board was 5×5 cm and the water tank was 30×30×20 cm. The probe was aligned to the center of the acrylic board and the water tank.

*B. Microbubble Trace Experiment*

To evaluate NSI-based PD performance in a controlled environment, we designed an experiment in which microbubbles descended away from the probe surface to the bottom of a beaker filled with degassed water. To enable this, the probe was first placed in degassed water containing microbubbles (Lantheus DEFINITY® microbubble), which adhered to the probe surface, and then the probe was placed in degassed water without microbubbles. This allowed observation of a small number of microbubbles that were initially attached to the surface of the probe and then pushed away by acoustic radiation force. An acoustic absorption material was placed at the bottom of the beaker to reduce reverberation. This setup is shown in Fig. 1.

The importance of the microbubble trace experiment relies on its capabilities of allowing us to observe the movement trace of a single microbubble, which is not feasible by using conventional flow phantom whose tube size is much larger than the diameter of a single microbubble. A single microbubble trace allows us to investigate the spatial resolution at a scale well below the diffraction limit.

The same ultrasound research platform, probe, and transmitted wave used for the element sensitivity measurement were used for the microbubble trace experiments. Nine plane waves were transmitted with steering angles from -4 degrees to 4 degrees in 1-degree increments. A post-compounding frame rate of 1,000 Hz was used in the data acquisition. In total, 1,600 frames with 9 angle acquisition were recorded for post-processing and generating the final image. In all the post-processing, the cutoff thresholds of SVD filter were adaptively chosen [5].

*C. In Vivo Experiment*

The protocol for the animal experiments was approved by the Institutional Animal Care and Use Committee (IACUC) at the University of Illinois at Urbana-Champaign. To evaluate the spatial resolution improvement for PD imaging when using NSI, we imaged the brain of a 10-week-old Sprague-Dawley rat. For all the experiments, the animal was anesthetized using isoflurane during the procedures.

Rat anesthesia was induced using a chamber supplying 4% isoflurane with oxygen. Once anesthesia was confirmed, the animal was placed on a stereotaxic frame with a nose cone supplied with 1.5% isoflurane for maintenance. The rat's head was secured in place with ear bars. After confirming that the animal was non-responsive to toe-pinch, the scalp was removed to expose the skull. A rotary micromotor with a 0.5 mm drill bit (Foredom K.1070, Bethel, CT) was used to open a cranial window approximately 1.2 cm x 0.9 cm, starting at bregma and centered along the sagittal suture. The ultrasound transducer was coupled directly to the surface of the brain with acoustic contact gel once the craniotomy was complete. An imaging plane approximately 2.5 mm caudal to bregma (β−2.5 mm) was selected for data acquisition.

For contrast injection, a 30-gauge catheter was inserted into the tail vein of the rat. Vessel patency was confirmed with a injection of 0.1 mL sterile saline. Lantheus DEFINITY® contrast agent was freshly activated, and then perfused through the tail vein catheter using a programmable syringe pump (NE-300, New Era Pump Systems Inc., Farmingdale, NY) at a rate of 50 uL/min.

The same scanning array probe and excitation procedures as III.B was used for the *in vivo* experiment.

*D. Image Quality Metrics*

To quantitatively evaluate the spatial resolution performance difference between NSI-based ultrafast PD microvessel imaging with that of conventional DAS, we evaluated the mean and standard deviation of the full width at half maximum (FWHM) of 20 manually selected microvessels in each *in vivo* data set. For NSI, the FWHM was evaluated for DC offset values ranging from 0.001 to 1 to exhibit the effects of the DC offset on the spatial resolution.

In addition to the FWHM estimates, which provide a local resolution performance evaluation, we also compared the resolution performance of the NSI-based ultrafast PD and conventional DAS-based PD in the Fourier domain [32]. The iso-frequency curves were plotted by calculating the mean value of the frequency component with the same spatial frequency. Then, an exponential curve fitting was applied on the iso-frequency curve before measuring the spatial frequency. The amplitude of the conventional DAS-based PD at half wavelength spatial frequency was marked and the same amplitude was used to find the corresponding spatial frequency of the NSI-based PD to determine the spatial resolution [33] for the microbubble trace experiment. The amplitude of the conventional DAS-based PD at one wavelength spatial frequency was marked and the same amplitude was used to find the corresponding spatial frequency of the NSI-based PD to determine the spatial resolution for the *in vivo* experiments.

*E. Computational Cost Measurement*

To evaluate the computational cost of the proposed method, we measured the processing time of the processing methods proposed in II.D and swept the lateral interpolation factor from 1 to 10 on a Nvidia RTX A6000 GPU.

IV. RESULTS

*A. Element Sensitivity Measurement*

The element sensitivity of the L22-14vX probe was measured for two transmit voltages: 6 V (the transmit voltage for microbubble trace experiments and contrast agent injected scan) and 30 V (the transmit voltage for contrast-free scan). The measured element sensitivity is plotted in Fig. 6. The microbubble trace data and contrast enhanced data were corrected for both the measurement at 6 V and the contrast-free data measured at 30 V to ensure the correction was accurate in case the element sensitivity was different at different transmit voltages.

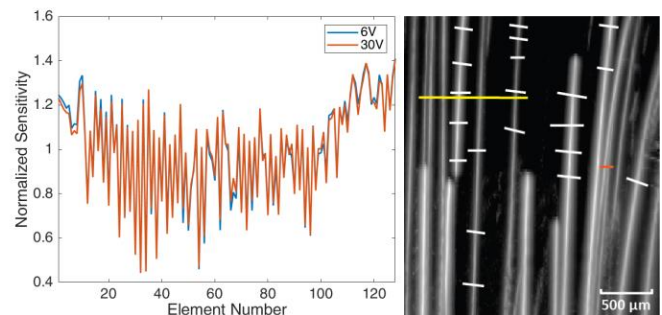

**Fig. 6. Measured element sensitivity of L22-14vX probe using transmit voltages of 6 V and 30 V (left). ROI selection for microbubble trace experiment (right). Yellow and orange solid lines: Cross-section profiles; White solid lines: FWHM measurement.**

*B. Microbubble Trace*

Both ESC and non-ESC PD microbubble trace images from both DAS and NSI using Pre-BF-SVD, Post-BF-SVD and Blockwise-SVD with DC offset = 0.1 are shown in Fig. 7. A movie that shows the microbubble movement is provided in supplemental video 1 to demonstrate the performance difference between DAS and NSI with ESC in a dynamic way.

The cross-sectional profiles of the microbubble traces for both DAS and NSI with and without ESC using three different SVD filters are shown in Fig. 8 and Fig. 9. The cross sections are marked with solid yellow and orange lines in Fig. 6. Twenty cross sections, which are marked with solid white lines in Fig.6, were selected to measure FWHM. The mean and standard deviation of the FWHM measured from the selected cross sections are shown in Fig.10 where the solid lines represent the mean FWHM, and the shaded regions represent the standard deviation of FWHM.

The Fourier domain images and the iso-frequency plots of both NSI with ESC and DAS using Post-BF-SVD with DC offset = 0.1 are shown in Fig.11. The global spatial resolution of NSI with ESC is 4.24 times better than DAS based on the exponential fit curve and the half wavelength amplitude cutoff.

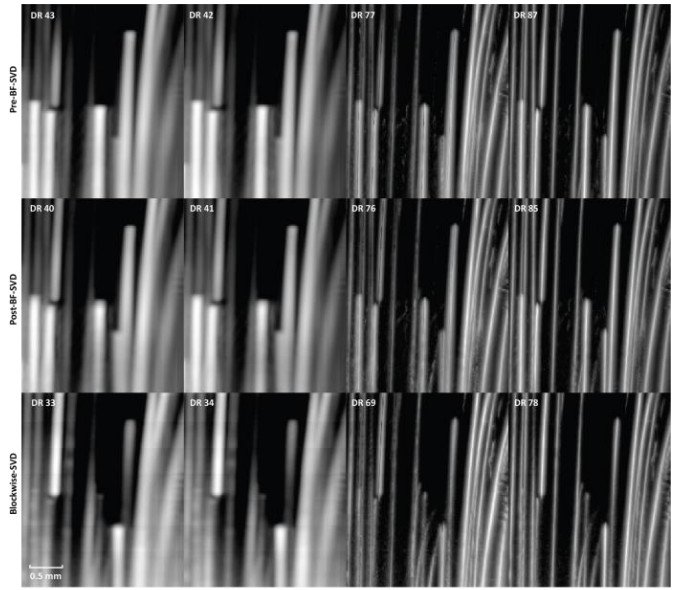

**Fig. 7.** PD images of microbubble trace. First column: DAS without ESC; Second column: DAS with ESC; Third column: NSI without ESC; Fourth column: NSI with ESC; First row: PD with Pre-BF-SVD clutter filter; Second row: PD with Post-BF-SVD clutter filter; Third row: PD with Blockwise-SVD filter.

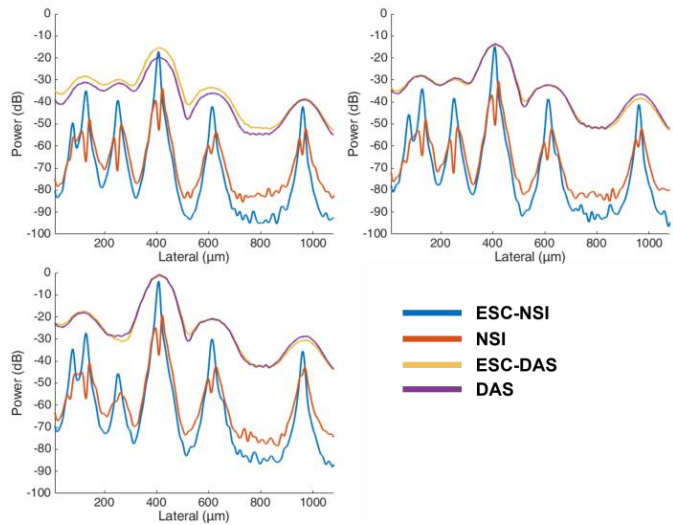

**Fig. 8.** Cross sectional profile of microbubble traces under different clutter filter settings Pre-BF-SVD (top left), Post-BF-SVD (top right) and Blockwise-SVD (bottom left) for both DAS and NSI with and without ESC with a DC offset of 0.1. The cross section is marked with a yellow solid line in Fig.6.

From Fig.8, we can observe that without ESC, NSI could generate two peaks instead of one peak for one microbubble trace which affects the resolution performance. From Fig. 10, we can observe that NSI could provide better spatial resolution with the ESC correction than NSI without ESC. The FWHM reduced more sharply with lower DC offset and ended up to a lower FWHM when using ESC. While for NSI without ESC, the FWHM did not decrease to the same level as NSI with ESC. For DAS, the presence of ESC did not demonstrate significant difference. Both the cross-sectional profile and the FWHM measurements demonstrate that ESC is necessary for NSI-based PD to reach its optimal spatial resolution without adding extra computational cost. In addition, the ESC does not affect the performance of DAS. Therefore, only NSI with ESC and DAS without ESC are evaluated in the following *in vivo* experiments.

From Fig. 9 we can observe that NSI allows us to distinguish two microbubble traces that are 25 μm apart from each other, which is only a quarter of the wavelength.

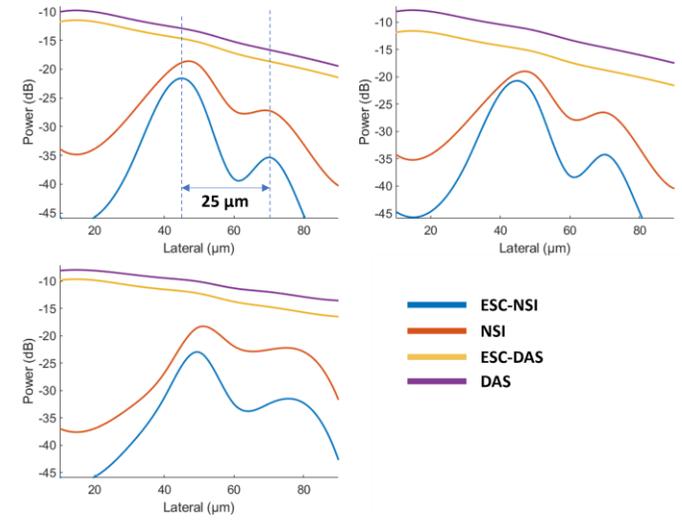

**Fig. 9.** Cross sectional profile of microbubble traces under different clutter filter settings Pre-BF-SVD (top left), Post-BF-SVD (top right) and Blockwise-SVD (bottom left) for both DAS and NSI using a DC offset of 0.1 with and without ESC. The cross section is marked with an orange solid line in Fig.6.

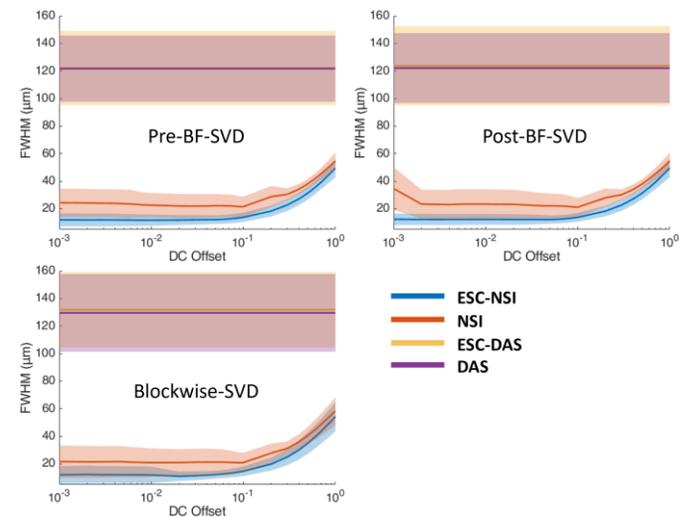

**Fig. 10.** FWHM profile of microbubble traces under different clutter filter settings Pre-BF-SVD (top left), Post-BF-SVD (top right) and Blockwise-SVD (bottom left) for both DAS and NSI with and without ESC. The FWHM measurement was taken from the 20 manually picked microbubble traces which are marked with white solid lines in Fig. 6. Note ESC-DAS and DAS lines were on top of each other.

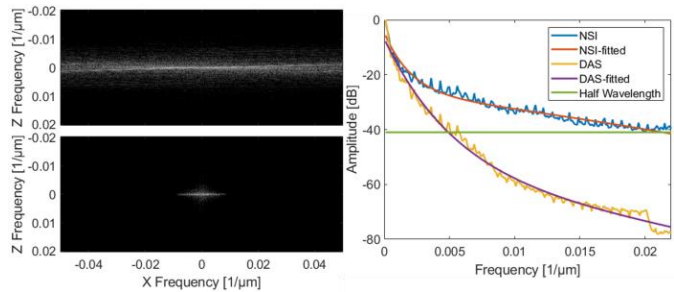

Fig. 11. Fourier domain image of NSI with ESC using Post-BF-SVD with DC offset = 0.1 (top left) and Fourier domain image of DAS using Post-BF-SVD (bottom left) and the iso-frequency plots (right) of both NSI with ESC and DAS without ESC using Post-BF-SVD with DC offset = 0.1.

### C. Rat Brain

PD microvessel images of the rat brain using contrast agents while imaging with DAS and NSI (with DC offset = 0.1) using Pre-BF-SVD, Post-BF-SVD and Blockwise-SVD are shown in Fig. 12. The zoomed-in version of Fig. 12 is shown in Fig. 13. The zoomed-in region is marked with a green box in Fig. 14. A movie that shows the microbubble movement is provided in supplemental video 2 to demonstrate the performance difference between DAS and NSI in a dynamic way.

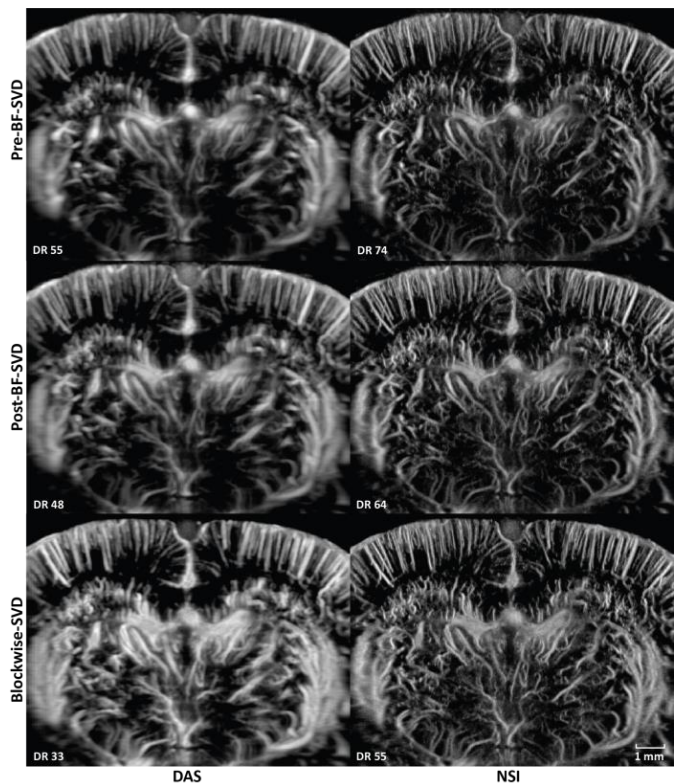

Fig. 12. PD images of contrast enhanced rat brain. First column: DAS; Second column: NSI; First row: PD with Pre-BF-SVD clutter filter; Second row: PD with Post-BF-SVD clutter filter; Third row: PD with Blockwise-SVD filter.

Three manually selected cross-sectional profiles of both NSI and DAS using three different SVD filters are shown in Fig. 15. These three cross sections are marked with solid color lines in Fig. 14. The cross-section that is marked with the blue solid line shows that NSI could resolve two vessels that are 39 μm separated. Similarly, the other two cross-sectional profiles also demonstrate that NSI could resolve vessels that could not be resolved by DAS.

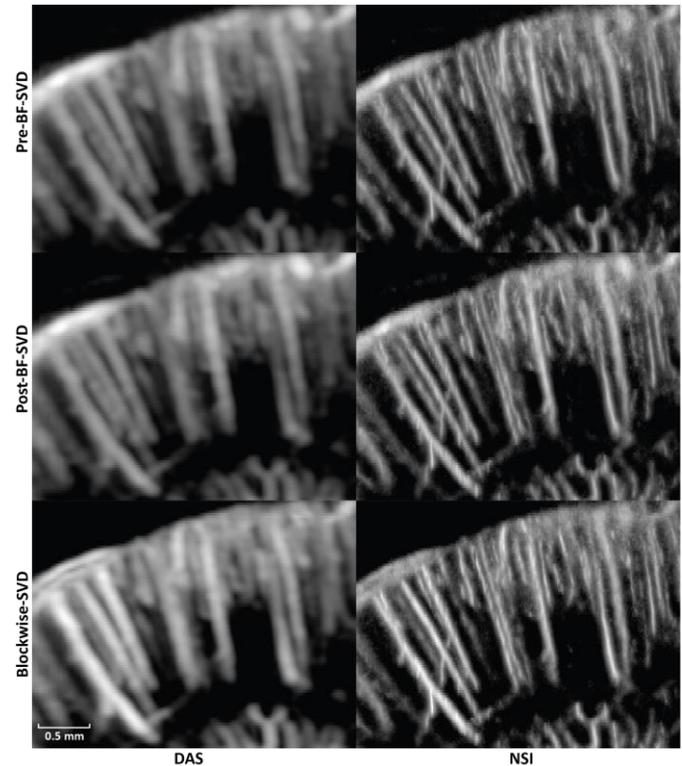

Fig. 13. Zoomed in PD images of contrast enhanced rat brain. The zoomed in region is marked with a green box in Fig. 14. First column: DAS; Second column: NSI; First row: PD with Pre-BF-SVD clutter filter; Second row: PD with Post-BF-SVD clutter filter; Third row: PD with Blockwise-SVD filter.

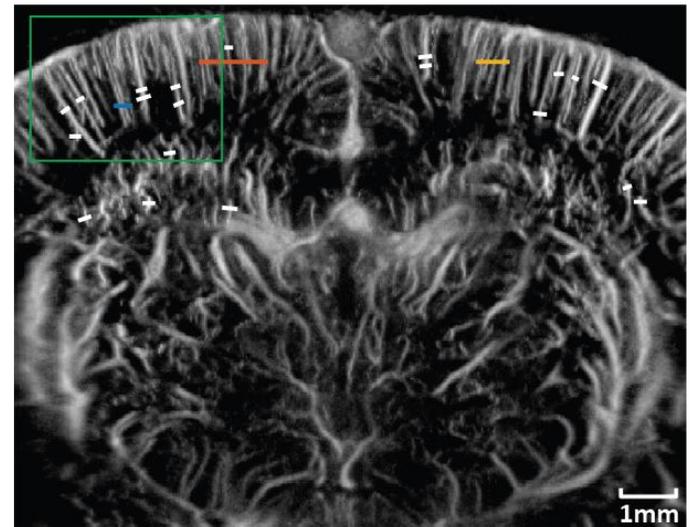

Fig. 14. ROI selection for contrast enhanced rat brain PD imaging. Blue solid line: First cross-section profile; Orange solid line: Second cross-section profile; Yellow solid line: Third cross-section profile; White solid lines: FWHM measurement. Green box: Zoomed in region.

Twenty cross sections which are marked with white solid lines in Fig. 14 were selected to measure FWHM. The mean and standard deviation of the FWHM measured from the selected cross sections are shown in Fig. 16 where the solid lines represent the mean FWHM, and the shadows represent the standard deviation of FWHM.

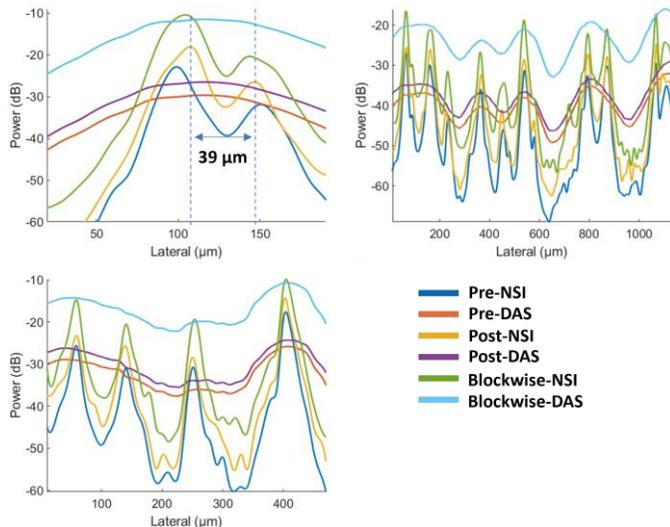

Fig. 15. Microvessel cross sectional profiles of PD microvessel images of contrast enhanced rat brain. Top left: first cross section marked with blue solid line in Fig.14; Top right: second cross section marked with orange solid line in Fig.14; Bottom left: third cross section marked with yellow solid line in Fig.14.

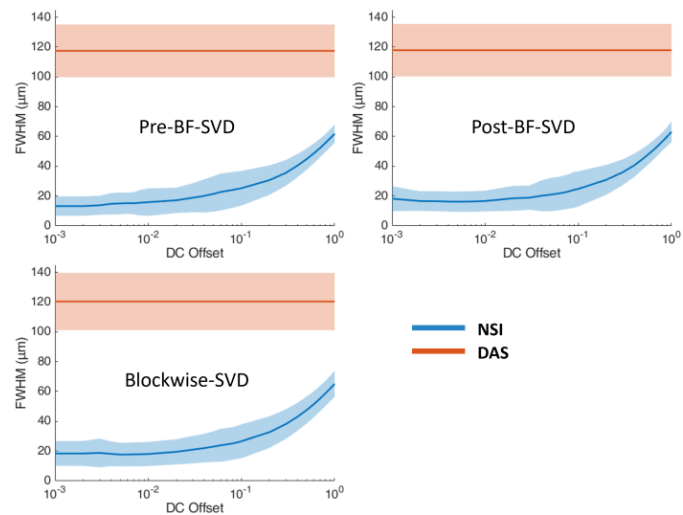

Fig. 16. FWHM profile of contrast enhanced rat brain vessels under different clutter filter settings Pre-BF-SVD (top left), Post-BF-SVD (top right) and Blockwise-SVD (bottom left) for both DAS and NSI. The FWHM measurement was taken from the 20 manually picked microbubble traces, which are marked with white solid lines in Fig. 14.

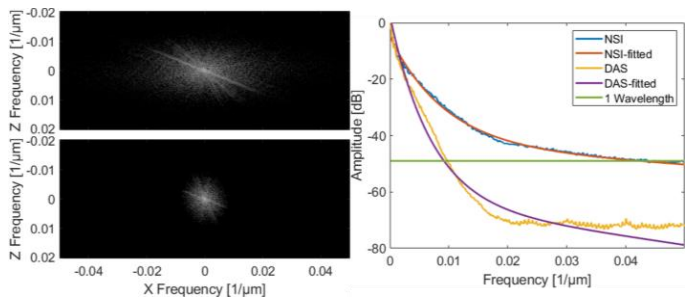

Fig. 17. Fourier domain image of NSI using Post-BF-SVD with DC offset = 0.1 (top left) and Fourier domain image of DAS using Post-BF-SVD (bottom left) and the iso-frequency plots (right) of both NSI (DC offset = 0.1) and DAS using Post-BF-SVD.

The Fourier domain images and the iso-frequency plots of both NSI and DAS using Post-BF-SVD with DC offset = 0.1 are shown in Fig. 17. The global spatial resolution of NSI is 4.25 times better than DAS based on the exponential fit curve and the one wavelength amplitude cutoff.

Following the same order of the contrast enhanced rat brain results and using the same DC offset for NSI, we present the contrast-free image in Fig.18 along with the zoomed in image in Fig. 19. The ROIs of different measurements are marked in Fig. 20. Supplemental video 3 demonstrates the performance difference between contrast-free DAS and NSI in a dynamic way.

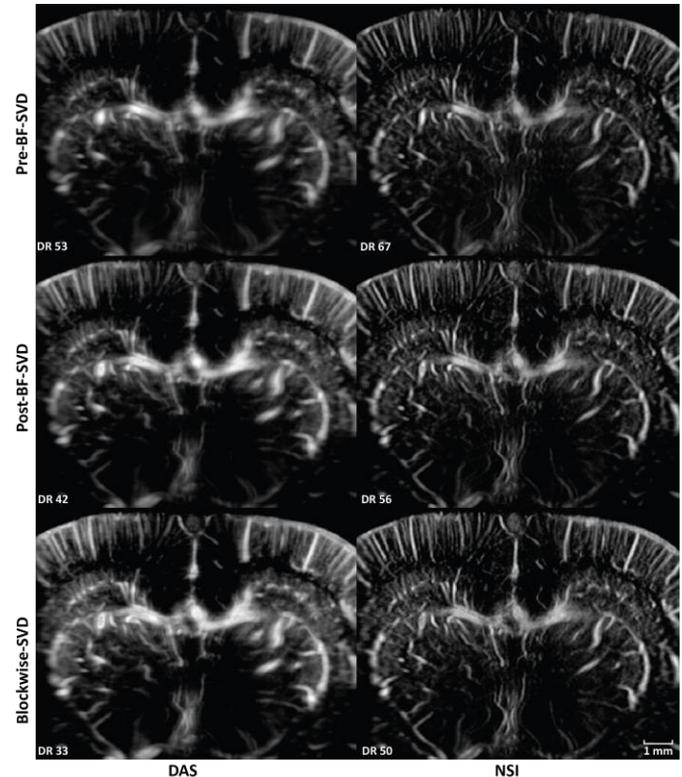

Fig. 18. PD images of contrast free rat brain. First column: DAS; Second column: NSI; First row: PD with Pre-BF-SVD clutter filter; Second row: PD with Post-BF-SVD clutter filter; Third row: PD with Blockwise-SVD filter.

Fig. 21 plots three manually selected cross-sectional profiles that are marked with solid color lines in Fig. 20. The cross-section profiles demonstrate that NSI could significantly narrow the width of vessels and enhance the contrast of the vessels from background. FWHM, which is measured from twenty manually selected cross sections is shown in Fig. 22.
Fig.23 shows the Fourier domain images and the iso-frequency plots. From the iso-frequency plots we measured that the global spatial resolution of NSI is 3.61 times better than DAS.

The Fourier domain images and the iso-frequency plots of both NSI (DC offset = 0.1) and DAS using Post-BF-SVD are shown in Fig. 22. The global spatial resolution of NSI is 3.61 times better than DAS based on the exponential fit curve and the one wavelength amplitude cutoff.

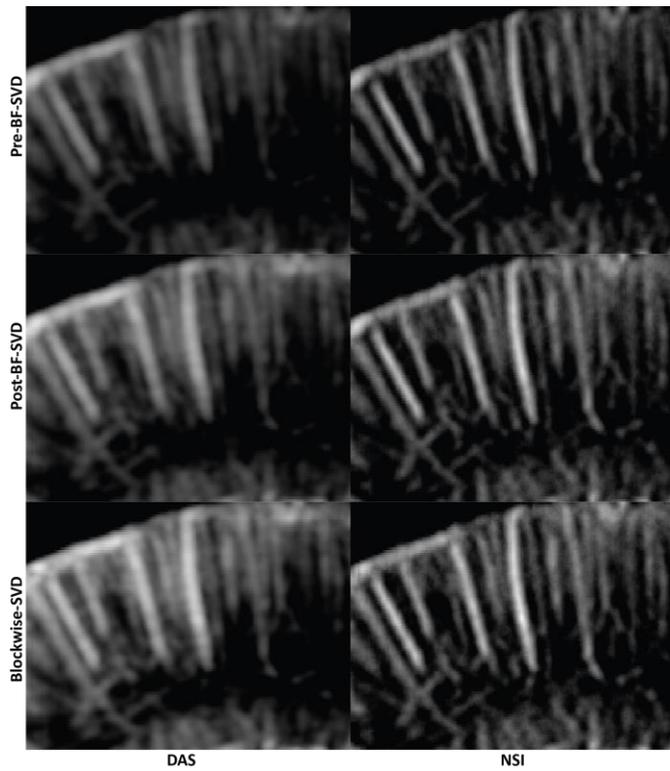

**Fig. 19.** Zoomed in PD images of contrast free rat brain. The zoomed in region is marked with a green box in Fig. 20. First column: DAS; Second column: NSI; First row: PD with Pre-BF-SVD clutter filter; Second row: PD with Post-BF-SVD clutter filter; Third row: PD with Blockwise-SVD filter.

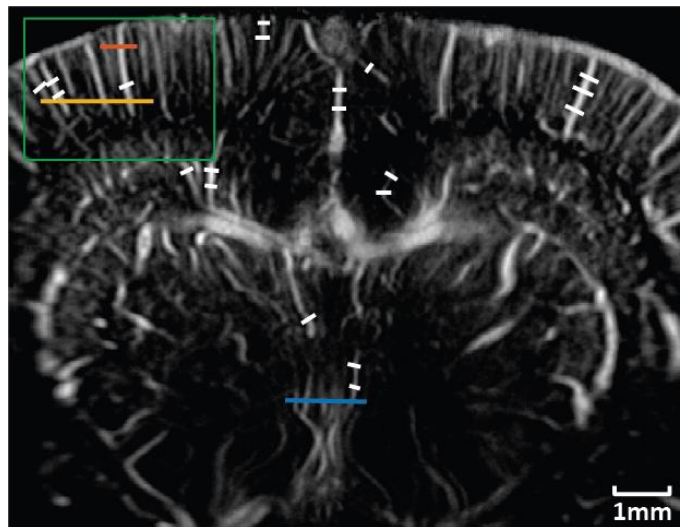

**Fig. 20.** ROI selection for contrast free rat brain PD imaging. Blue solid line: First cross-section profile; Orange solid line: Second cross-section profile; Yellow solid line: Third cross-section profile; White solid lines: FWHM measurement. Green box: Zoomed in region.

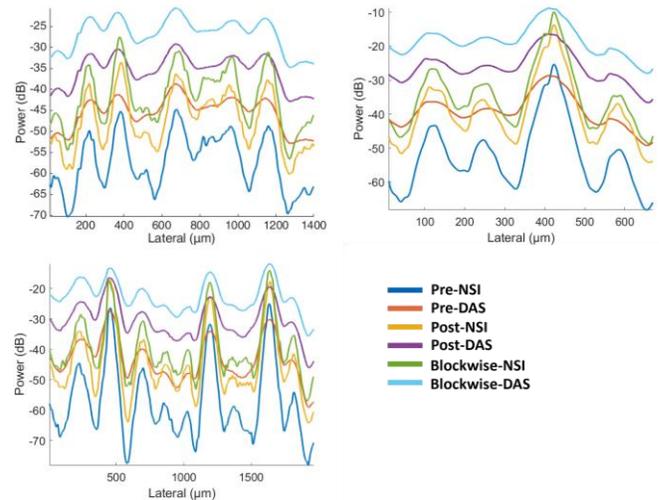

**Fig. 21.** Microvessel cross sectional profiles of PD microvessel images of contrast free rat brain. Top left: first cross section marked with blue solid line in Fig. 20; Top right: second cross section marked with orange solid line in Fig. 20; Bottom left: third cross section marked with yellow solid line in Fig. 20.

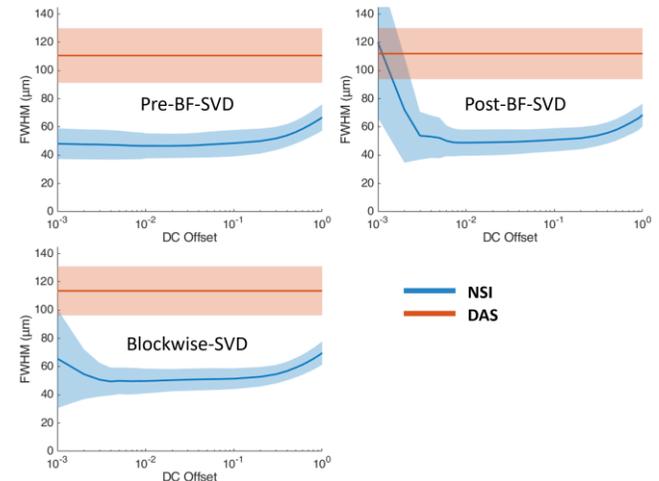

**Fig. 22.** FWHM profile of contrast-free rat brain vessels under different clutter filter settings Pre-BF-SVD (top left), Post-BF-SVD (top right) and Blockwise-SVD (bottom left) for both DAS and NSI. The FWHM measurement was taken from the 20 manually picked microbubble traces which are marked with solid white lines in Fig.19.

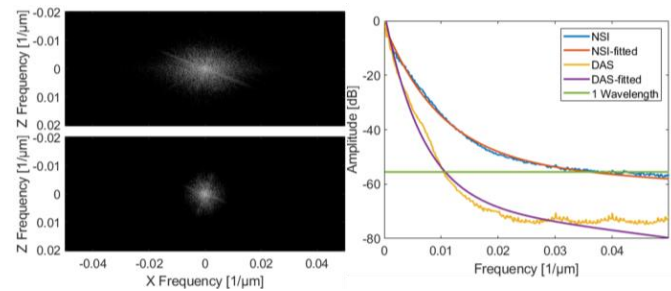

**Fig. 23.** Fourier domain image of NSI using Post-BF-SVD with DC offset = 0.1 (top left) and Fourier domain image of DAS using Post-BF-SVD (bottom left) and the iso-frequency plots (right) of both NSI (DC offset = 0.1) and DAS using Post-BF-SVD.

*D. Computational Cost*

The computational cost which is measured by the duration of the computation time is shown in Fig. 24. At low interpolation factors, the Post-BF-SVD has lower computation time compared with the Pre-BF-SVD. At high interpolation factors, the Pre-BF-SVD has lower computation time compared with Post-BF-SVD. This is because the higher the interpolation factor, the longer the SVD computation time of the Post-BF-SVD clutter filter because Post-BF-SVD is operated on the beamformed IQ data whose size increases as the interpolation factor increases. For the Pre-BF-SVD clutter filter, the interpolation factor does not affect the SVD computation time as the interpolation is performed during the beamforming, which happens after the SVD clutter filter. The Blockwise-SVD has the highest computational cost as hundreds of SVD computations are performed. NSI always has around a 40% longer computational time than DAS with three different SVD filters and different interpolation factors.

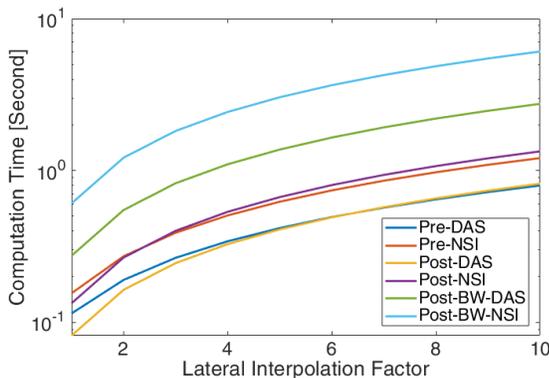

**Fig. 24. Measured computation time per frame with respect to the lateral interpolation factor with six different processing methods.**

## V. Discussion

In this study, NSI-based ultrafast PD microvessel images were compared with DAS-based images using three different SVD clutter filtering techniques in the microbubble trace experiment to explore the resolution performance improvements of the proposed methods to the limit of single microbubble size. Next, *in vivo* experiments were performed by scanning a rat brain with and without the presence of contrast agents.

The image quality was first visually compared between NSI-based and DAS-based PD images in both microbubble trace experiment and *in vivo* experiments using three different SVD clutter filters with adaptive DR to illustrate the resolution differences under the same vessel to background contrast. It was also compared using manually selected cross-sectional profiles to provide a closer look at the spatial resolution and contrast improvements of NSI-based PD. The spatial resolution was compared using different quantitative metrics such as FWHM and the spatial frequency cutoff of same amplitude on the iso-frequency curve. NSI improved spatial resolution in the following aspects.

First, the improvement of spatial resolution from NSI resulted in observing microvessels close to each other that were unresolved in DAS. For example, in Fig. 9, two microbubble traces that were 25 µm away from each other could be resolved by NSI, and in Fig. 15, two microvessels that were separated by 39 µm were not resolvable in the DAS images, while they were clearly resolved in NSI images. This demonstrates that the imaging performance gained from NSI yields improved spatial resolution of the underlying vascular physiology and is not merely due to extending the dynamic range of the PD image.

The mean FWHM results also support improvement in the NSI images. The mean FWHM was reduced from 120 µm using DAS to 20 µm using NSI with contrast agents and 110 µm with DAS to 50 µm using NSI without contrast agents. NSI-based PD also provides smaller variance in the FWHM which suggests that NSI-based PD provides a more homogeneous spatial resolution performance. Given the frequency of the ultrasound pulse and the F-number of the beamformer (i.e., constant F-number of 1), the theoretical resolution limit is ~100 µm. DAS-based PD provided similar performance to this theoretical resolution limit. The NSI-based PD overcame this theoretical resolution limit by a factor of five with the presence of contrast agents and a factor of two without the presence of contrast agents. The spatial resolution of NSI depends on the SNR, which was increased with the contrast agents allowing for a lower DC offset value to be used.

Second, in terms of global resolution, over 4 times higher spatial frequency cutoff of the same amplitude on the iso-frequency curve has been measured from the NSI-based PD compared to that of the DAS-based PD with the presence of contrast agents. Without contrast agents, NSI-based PD could still provide over 3 times higher spatial frequency cutoff than that of DAS-based PD. This trend coincides with the previous local mean FWHM measurements.

Third, the importance of ESC to NSI has been demonstrated in microbubble trace experiments. The results indicate that ESC could significantly improve the NSI-based PD performance especially from the cross-sectional profile plot. From Fig. 8, we can observe that without ESC, NSI-based PD could not properly detect the microbubble trace. From Fig. 10, we can observe that the FWHM of NSI without ESC was larger than that of NSI with ESC. As ESC is a correction method without adding extra computational cost and easy to implement, we suggest the use of ESC to implement NSI when pushing the spatial resolution to low DC offsets. Though the measurement of ESC could cost extra time and setup, two ESC measurements could be separated by weeks or months depending on the speed of element deterioration.

Furthermore, the image quality was related to the choice of DC offset value. We swept the DC offset value from 0.001 to 1 for all the experiments. The results suggest that the smaller the DC offset the better the spatial resolution, when DC offset is in the range between 1 and 0.1. When DC offset is smaller than 0.1, the spatial resolution improvement starts to saturate. When the DC offset is approaching 0.001, the resolution performance starts to break down as the noise could dominate the signal, especially for the contrast-free, *in vivo* experiment. As a result, we suggest choosing DC offset at 0.1 as an empirical choice for NSI-based PD imaging. Potentially, some denoising methods could be performed on unbeamformed RF data to overcome this limit for NSI and further improve the contrast-free PD microvessel imaging performance.

The traditional quantitative metrics of image quality, such as SNR and CNR, were not included in this study as recent studies have shown that these measurements could be adjusted by dynamic range alterations [34][35][36][37], and they are highly dependent on the ROI selection. To provide a fairer comparison, the images were rendered with adaptive DR, which is determined by the statistics of the log compressed power Doppler images. The adaptive DR allowed us to view the PD images at the same contrast and brightness between vessel and background for the different imaging methods. From the automatically generated adaptive DR values, we observed that NSI-based images had a much larger variance compared with that of DAS-based PD images. Though NSI-based PD images were shown with a much higher DR to keep both DAS-based PD and NSI-based PD at the same contrast and brightness, they still demonstrated a much higher spatial resolution, which can be observed from the cross-sectional profile plotted in Fig. 15 and better detection in small vessels which is shown in Fig. 19. In addition, we did not observe the existence of any extra clutter from NSI-based PD images, which suggests that NSI did not result in a higher clutter level.

From the computational cost and data acquisition cost aspect, the NSI-based ultrafast PD does not require a longer acquisition period and increase the computational time to a higher order compared to DAS-based ultrafast PD. The same set of data for DAS-based ultrafast PD can be used for NSI-based ultrafast PD as the NSI-based ultrafast PD is a fully post-processing method, which does not require any special transmit and receive sequence. From Fig. 24 we can observe that the computational time for NSI-based ultrafast PD was only 40% higher than that of DAS-based ultrafast PD. At the same time, NSI-based ultrafast PD provided a four-fold or higher resolution improvement, which are shown in Figs. 11, 17 and 23. As a result, any clinical machines with plane wave imaging capability could significantly improve the spatial resolution by enabling NSI-based ultrafast PD with an upgrade in their software based beamformer or GPU beamformer.

In terms of different SVD filters, both Pre-BF-SVD and Post-BF SVD filters demonstrate similar imaging performance and computational cost for NSI-based PD. When the lateral interpolation factor is small, the Post-BF SVD filters have lower computational cost and when the lateral interpolation factor is large, the Pre-BF SVD filters have lower computational cost. The Blockwise-SVD provides a better image quality in the deeper region with a much higher computational cost.

The contrast-enhanced and contrast-free *in vivo* experiments demonstrate that NSI could not only improve the spatial resolution with the occurrence of contrast agents, but also without the occurrence of contrast agents. However, NSI does not provide the same improvements in both cases which indicates that the spatial resolution performance of NSI is still limited by the SNR of the input signal as the contrast agents could provide a much higher SNR than the native blood flow.

## VI. Conclusion

In this work, a framework of beamforming-based high resolution ultrafast PD imaging has been proposed and evaluated with both microbubble trace and *in vivo* datasets and compared with DAS-based PD imaging. The results demonstrate the potential to overcome the diffraction limit without the need for localization and tracking. The spatial resolution was both qualitatively and quantitatively compared between NSI- and DAS-based images. Results suggest that the NSI-based PD technique provided four-fold or higher spatial resolution improvements over DAS with only 40% increase in computation time.